%% file: interfaces.tex
\documentclass[letterpaper,twocolumn,10pt]{article}
\usepackage{usenix2019_SOUPS}

\usepackage{booktabs} 
\usepackage{multirow}
\usepackage{balance}
\usepackage{graphicx}

\usepackage[skip=0.5\baselineskip]{caption}

\usepackage[toc,page]{appendix}
\usepackage{latexsym}
\usepackage[noend]{algpseudocode}
\usepackage{enumitem}

\usepackage{tabu}

\newenvironment{Quote}%
{\begin{list}{}%
        {\setlength\leftmargin{\parindent}\setlength\rightmargin{\leftmargin}}%
\item[]\makebox[0pt][r]{``}\ignorespaces}
{\unskip\makebox[0pt][l]{''}
\end{list}}

\begin{document}

\date{}
\title{\Large \bf System Administrators Prefer Command Line Interfaces, Don't They?\\An Exploratory Study of Firewall Interfaces\thanks{To appear in the proceeding of the fifteenth Symposium on Usable Privacy and Security (SOUPS 2019)}}

\def\plainauthor{Artem Voronkov, Leonardo A. Martucci, Stefan Lindskog}

\author{
	{\rm Artem Voronkov}\\
	Karlstad University
	\and
	{\rm Leonardo A. Martucci}\\
	Karlstad University
	\and
	{\rm Stefan Lindskog}\\
	Karlstad University
}

\maketitle
\pagestyle{empty}

\input{./tex/abstract}

\input{./tex/introduction}

\input{"./tex/related_work"}

\input{./tex/methodology}

\input{./tex/results}

\input{./tex/discussion}

\input{./tex/conclusion}

\input{./tex/acknowledgment}
\input{./tex/availability}

\balance
\bibliographystyle{plain}
\bibliography{interfaces}

\renewcommand\appendixpagename{Appendix}
\input{./tex/appendix}

\end{document}

%% file: tex/abstract.tex
\begin{abstract}
A graphical user interface (GUI) represents the most common option for interacting with computer systems. However, according to the literature system administrators often favor command line interfaces (CLIs). The goal of our work is to investigate which interfaces system administrators prefer, and which they actually utilize in their daily tasks. We collected experiences and opinions from 300 system administrators with the help of an online survey. All our respondents are system administrators, who work or have worked with firewalls. Our results show that only $32$\% of the respondents prefer CLIs for managing firewalls, while the corresponding figure is $60$\% for GUIs. We report the mentioned strengths and limitations of each interface and the tasks for which they are utilized by the system administrators. Based on these results, we provide design recommendations for firewall interfaces.
\end{abstract}

%% file: tex/introduction.tex
\section{Introduction}
\label{sec:introduction}

Firewalls are systems designed to regulate network traffic, and are often the first line of defense in computer networks. The maintenance and configuration of firewalls is the responsibility of system administrators. System administrators have multiple methods available to interact with firewalls, e.g. via a command line interface (CLI), graphical user interface (GUI), or application programming interface (API). Although visualization offers an effective approach to exploring and managing data, the use of GUIs by system administrators is not taken for granted. According to the literature, the main instrument for system administrators is the CLI \cite{takayama2006trust, Botta07, haber2007design}.

In this paper, we examine how system administrators interact with firewalls. The goal of our study is to gain a better understanding of the following questions: 
\begin{itemize}
\item[Q1:] What firewall interfaces do system administrators use? 
\item[Q2:] What firewall interfaces do they prefer?
\end{itemize}
Additionally, we want to gain insights into which of the interfaces are beneficial for which tasks, and what strengths and limitations they have. To answer our research questions, we surveyed 300 system administrators and collected their experiences and opinions of utilized firewall interfaces through an online survey.

Unexpectedly, our results show that $70$\% of the system administrators work primarily with firewall GUIs, with $60$\% preferring GUIs as a main instrument. The system administrators mainly choose GUIs because they provide better visual representations of data, are easier to create and modify rules with, and are convenient for occasional use.
Relatively few system administrators utilize a CLI as their primary or preferred firewall interface: $24$\% and $32$\%, respectively. According to our respondents, the main reasons for choosing command line interfaces are their flexibility, efficiency of use, superior functionality, and performance; aspects in which GUIs are deficient. 

The contributions of our work are summarized as follows:
\begin{itemize}
\item We conduct an online study on the preferences of system administrators regarding firewall interfaces, with 300 volunteer participants.
\item Using the gathered data, we classify and report the main strengths and limitations of CLIs and GUIs.
\item We provide insights into tasks in which utilizing a CLI or GUI is advantageous for system administrators.
\item We provide some recommendations for designers and developers of firewall interfaces, taking into account the main problems of the two interfaces.
\end{itemize}

The remainder of this paper presents a review of work related to our study in Section \ref{sec:rw}, describes our research methodology in Section \ref{sec:methodology}, and presents the results in Section \ref{sec:results}. A discussion of the findings, limitations, and our design recommendations is presented in Section \ref{sec:discussion}. Finally, concluding remarks are provided in Section \ref{sec:conclusion}.

%% file: tex/related_work.tex
\section{Related Work}\label{sec:rw}

Despite the fact that GUIs are known to be convenient for the presentation of large amounts of information, their use is limited in the field of system configuration, as noted by Mahendiran et al. \cite{mahendiran2014exploring}.

Botta et al.\ \cite{Botta07} and Haber and Bailey \cite{haber2007design} reported the results of two independent ethnographic studies describing the routines and activities of system administrators. Haber and Bailey followed the daily work of three system administrators, and reported their preference of CLIs over GUIs owing to their speed, scalability, reliability, transparency and trustworthiness. These findings are in line with the interviews of Botta et al.,\ involving a dozen {\small IT} professionals who reported being more comfortable with CLIs than GUIs, especially because of their versatility. Botta et al.\ also highlighted the reliability problem of GUIs that \emph{``write configuration files that sometimes do not take effect''} and \emph{``write unnecessary, noisy markup into configuration files.''}  

For a study with 101 participants, Takayama and Kandogan \cite{takayama2006trust} reported that $65$\% of the participants were primarily CLI users, because CLIs are considered to be more reliable, fast, robust, trustworthy, and accurate. Furthermore, the authors pointed out that trust is critical in the adoption of a technology.

However, system administrators require graphical tools that can facilitate their daily work and make it less error-prone \cite{mahendiran2014exploring}. This is especially relevant for security system administrators, as their work has been demonstrated to be more complex \cite{gagne2008identifying}. 

Recent research has sought to leverage the benefits of information visualization in designing interfaces for network security. Shiravi et al. \cite{shiravi2012survey} presented a survey of visualization systems in network security in general, while Voronkov et al. \cite{voronkov2018systematic} reviewed papers specifically concerning firewalls. The authors of both papers identified limitations of existing visualization techniques and suggested future research directions.

Xu et al. \cite{xu2016hci} argued that \emph{``system configuration becomes a new human--computer interaction (HCI) problem,''} and that \emph{``classic interface design principles are not sufficient for system configuration.''} A variety of research studies \cite{haber2007design,  thompson2007command, velasquez2008designing} have attempted to address these problems and suggest appropriate design principles for system configuration.

Although interface preferences of system administrators have been studied in the literature, the present work represents the first large-scale study investigating firewall interfaces, with $300$ participants. Furthermore, we aim to investigate whether there have been changes in preferences, as it has been over 10 years since the studies of Botta et al.\ \cite{Botta07}, Haber and Bailey \cite{haber2007design}, and Takayama and Kandogan \cite{takayama2006trust} were published. Another important aspect of our work is the qualitative analysis of participants' comments regarding the strengths and limitations of firewall interfaces, as well as tasks in which these interfaces are superior.

%% file: tex/methodology.tex
\section{Methodology}
\label{sec:methodology}

We collected both quantitative and qualitative data on the interactions between system administrators and firewall interfaces through an online survey ($N=300$). In this section, the methodology and demographics of the participants are described, while the remainder of the quantitative data and qualitative results are presented in Section \ref{sec:results}.

\subsection{Survey Details}

We collected the data through an online survey, which ran for six weeks from April to June 2018.\footnote{The survey is available at \url{https://www.soscisurvey.de/firewall_interfaces/}}
The survey utilized {skip logic} (also known as {branch logic} or {conditional branching}) and consisted of up to 14 questions, four of which were open-ended. The close-ended questions required an answer and we also encouraged the participants to answer the open-ended questions, although these were not mandatory. 

The survey consisted of two parts. In the first, we asked the participants about the following aspects of their interactions with firewalls:
\begin{itemize}
\item How much time on average they spend working with firewalls.
\item Which firewall interface they mainly work with, and which interface they prefer.
\item Which tasks are easier with which firewall interface.
\item What strengths and limitations those interfaces have.
\end{itemize}

Only general questions about firewall interfaces were asked in the survey. No questions about specific vendor solutions were included. In the second part of the survey, demographically related questions were asked, such as on age, gender, and expertise.

We kept the survey short to minimize respondent fatigue. The survey took an average of 177 seconds ($SD=106$, \mbox{$M=148$}, $Q1=101$, and $Q3=228$ seconds) of the participants' time to be answered.

Prior to dissemination, the survey was pre-tested with six users. Based on their feedback, a few questions were slightly altered to eliminate some ambiguity in the wording, although no significant changes were necessary. For wider coverage, the survey was translated from the original (English) language into three others (Portuguese, Russian, and Swedish) by bilingual speakers.

\subsection{Recruitment and Participants}

The participants for the study were recruited using various channels:
\begin{enumerate}
\item System administrators' forums. The ``Sysadmin'' subreddit yielded the majority of our participants.\footnote{\url{https://www.reddit.com/r/sysadmin/}} Another contributor was the SysAdmins.ru forum.\footnote{\url{https://sysadmins.ru/}}
\item System administrators' mailing lists. We contacted several system administrators from our professional networks and asked them to distribute the survey via system administrator mailing lists of which they are members. 
\end{enumerate}

Of 516 participants that started our online survey, 303 completed it (ca. 59\% completion rate). After the quality check, three participants were removed as they filled out nonsensical answers. Table \ref{tab:demographics} summarizes the demographics of the remaining 300 participants. Our sample is heavily skewed owing to specificity of the target audience (the percentage of female system administrators is known to be very low \cite{ashcraft2016women}) and recruitment method. A majority of the participants (approximately 80\%) were recruited via the ``Sysadmin'' subreddit, which led to the sample being more male (only 7.5\% of the subreddit members are female \cite{burkhart2017subreddit}) and younger than the general population, owing to the demographics of Reddit users \cite{sattelberg2018demographics}. All participants were volunteers, and no financial compensation was offered.

\begin{table}[t]
\caption{Participant demographics ($N=300$).}
\label{tab:demographics}
\resizebox{\columnwidth}{!}{%
\begin{tabular}{lll}
\hline
 & \textbf{Metric} & \textbf{Participants} \\ \hline
\multirow{6}{*}{\textbf{Age}} & 18-24 & 34 (11.3\%) \\
 & 25-34 & 142 (47.3\%) \\
 & 35-44 & 86 (28.7\%) \\
 & 45-54 & 25 (8.3\%) \\
 & 55-64 & 9 (3.0\%) \\
 & Prefer not to answer & 4 (1.3\%) \\
 &  &  \\
\multirow{4}{*}{\textbf{Gender}} & Female & 3 (1.0\%) \\
 & Male & 285 (95.0\%) \\
 & Other & 1 (0.3\%) \\
 & Prefer not to answer & 11 (3.7\%) \\
 &  &  \\
\multirow{6}{*}{\textbf{\begin{tabular}[c]{@{}l@{}}Time per week\\ (on average) \\spent on\\ managing\\ firewalls\end{tabular}}} & \textless{}1 hour/week & 106 (35.3\%) \\
 & 1-4 hours/week & 117 (39.0\%) \\
 & 5-8 hours/week & 35 (11.7\%) \\
 & 9-12 hours/week & 11 (3.7\%) \\
 & 13+ hours/week & 21 (7.0\%) \\
 & Do not directly manage firewalls & 10 (3.3\%) \\
 &  &  \\
\multirow{5}{*}{\textbf{\begin{tabular}[c]{@{}l@{}}Experience as\\ system\\ administrator\end{tabular}}} & \textless{}1 year & 6 (2.0\%) \\
 & 1-3 years & 46 (15.3\%) \\
 & 4-6 years & 64 (21.3\%) \\
 & 7-9 years & 39 (13.0\%) \\
 & 10+ years & 145 (48.3\%) \\
 &  &  \\
\multirow{4}{*}{\textbf{\begin{tabular}[c]{@{}l@{}}Proficiency\\ with \\ firewalls\end{tabular}}} & Basic knowledge & 20 (6.7\%) \\
 & Intermediate & 114 (38.0\%) \\
 & Advanced & 114 (38.0\%) \\
 & Expert & 52 (17.3\%) \\
 &  &  \\
\multirow{4}{*}{\textbf{Language}} & English & 256 (85.3\%) \\
 & Portuguese & 7 (2.3\%) \\
 & Russian & 21 (7.0\%) \\
 & Swedish & 16 (5.3\%) \\ \hline
\end{tabular}
}
\end{table}

\subsection{Survey Data Analysis}
The data were analyzed using a content analysis approach. With this approach, it is possible to analyze data qualitatively at the same time as quantifying it \cite{grbich2012qualitative}. 

Two of the authors worked independently and coded participants' responses to the open-ended questions using an initial (open) coding approach \cite{saldana2012coding}. Two coding procedures were performed: one before and one after the final codebook. We utilized NVivo for all coding.\footnote{\url{https://www.qsrinternational.com/nvivo/home}} NVivo helped us to organize and analyze the qualitative data, i.e. open-ended survey responses. NVivo provides methods to automatically or manually code the data. We used manual coding only, which comprises three approaches: 1) select and code content, 2) drag and drop selected content, and 3) in vivo coding.

After the authors completed the first coding procedure, they met, discussed their codes, consolidated them, and formed a final codebook, which consisted of 230 codes (see Section~\ref{sec:availability}).
Using the final codebook during the second coding procedure, 1570 coding references were identified. It is worth mentioning that each answer from a participant can have several different codes associated with it, but at most one instance of a single code.

The Cohen's kappa inter-rater reliability value for the final codes was $0.79$, indicating an excellent agreement between the \mbox{coders \cite{fleiss2003statistical}}. The cases in which the coders varied in the final codes were resolved by the first author, who examined respondents' answers and assigned the most appropriate code.

\subsection{Ethical Considerations}
The survey was conducted in accordance with the Swedish Ethical Review Act \cite{etik2003} and the Good Research Practice guidelines from the Swedish Research Council \cite{etik2018}. No sensitive personal data were collected and no mental or physical interventions took place. Therefore, no explicit ethical approval was required for this study. The following precautions were taken into consideration to ensure that the participants were treated ethically and with respect:
\begin{itemize}
\item The participants provided informed consent before starting the survey. The informed consent form stated the purpose of the study, its approximate duration, our commitment to confidentiality, and their rights as participants, including the right to withdraw from the study at any point in time.
\item Only (the minimal) necessary personal data (see Table~ \ref{tab:demographics}) were collected.  
\item No sensitive personal data were collected.
\end{itemize}

%% file: tex/results.tex
\section{Results}\label{sec:results}

We describe the survey results by providing both quantitative and qualitative data in Sections \ref{subsec:quan}--\ref{subsec:qual}. In Section~ \ref{subsec:suit} we report on the suitability of firewall CLIs and GUIs for different tasks.

\begin{table*}[t]
\centering
\caption{Relations between primary and preferred firewall interfaces based on the answers from our survey.}\label{tab:pripre}
\begin{tabular}{|p{1.94cm}l|c|c|c|c|l|}
\hline
 &  & \multicolumn{4}{c|}{\textbf{Preferred interface}} &  \\ \cline{3-6}
 &  & \begin{tabular}[c]{@{}l@{}}CLI\end{tabular} & \begin{tabular}[c]{@{}l@{}}GUI\end{tabular} & \begin{tabular}[c]{@{}l@{}}API\end{tabular} & Other & \textbf{Total} \\ \hline
\multicolumn{1}{|l|}{\multirow{4}{*}{\begin{tabular}[c]{@{}l@{}}\textbf{Primary}\\ \textbf{interface}\end{tabular}}} & \begin{tabular}[c]{@{}l@{}}CLI\end{tabular} & 61 & 7 & 3 & 2 & 73 (24.3\%) \\ \cline{2-7} 
\multicolumn{1}{|l|}{} & \begin{tabular}[c]{@{}l@{}}GUI\end{tabular} & 30 & 169 & 4 & 6 & 209 (69.7\%) \\ \cline{2-7} 
\multicolumn{1}{|l|}{} & \begin{tabular}[c]{@{}l@{}}API\end{tabular} & 0 & 1 & 4 & 0 & 5 (1.7\%) \\ \cline{2-7} 
\multicolumn{1}{|l|}{} & Other & 4 & 2 & 0 & 7 & 13 (4.3\%) \\ \hline
\textbf{Total} &  & 95 (31.7\%) & 179 (59.7\%) & 11 (3.6\%) & 15 (5.0\%) & 300 (100.0\%) \\ \hline
\end{tabular}
\end{table*}

\begin{table*}[]
\centering
\caption{Relations between firewall proficiency and preferred firewall interfaces based on the answers from our survey.}\label{tab:profpref}
\begin{tabular}{|p{1cm}l|c|c|c|c|l|}
\hline
 &  & \multicolumn{4}{c|}{\textbf{Preferred interface}} &  \\ \cline{3-6}
 &  & \begin{tabular}[c]{@{}l@{}}CLI\end{tabular} & \begin{tabular}[c]{@{}l@{}}GUI\end{tabular} & \begin{tabular}[c]{@{}l@{}}API\end{tabular} & Other & \textbf{Total} \\ \hline
\multicolumn{1}{|l|}{\multirow{4}{*}{\textbf{Proficiency}}} & Basic knowledge & 5 & 14 & 0 & 1 & 20 (6.7\%) \\ \cline{2-7} 
\multicolumn{1}{|l|}{} & Intermediate & 30 & 72 & 2 & 10 & 114 (38\%) \\ \cline{2-7} 
\multicolumn{1}{|l|}{} & Advanced & 43 & 65 & 5 & 1 & 114 (38\%) \\ \cline{2-7} 
\multicolumn{1}{|l|}{} & Expert & 17 & 28 & 4 & 3 & 52 (17.3\%) \\ \hline
\textbf{Total} &  & 95 (31.7\%) & 179 (59.7\%) & 11 (3.6\%) & 15 (5.0\%) & 300 (100.0\%) \\ \hline
\end{tabular}
\end{table*}

\subsection{Quantitative Data}\label{subsec:quan}

Seventy percent of the participants in our survey are primarily firewall GUI users, and $60$\% prefer GUIs to text-based interfaces when having to deal with a firewall (see \mbox{Table \ref{tab:pripre}}). Approximately a quarter of the polled system administrators primarily work with textual interfaces ($24$\% for CLI and $2$\% for API), and slightly over one third prefer to use these as their main interface: $32$\% and $4$\% for CLIs and APIs, respectively.
The option \texttt{Other} indicates system administrators that use either a combination of the aforementioned interfaces or another type of firewall interface.

Based on our data, there may be a connection between a system administrator's proficiency with firewalls and the interface that they prefer to utilize. 
Table \ref{tab:profpref} shows that the stronger the firewall expertise of respondents, the lower the likelihood of utilizing GUIs. Seventy percent of the system administrators with a basic knowledge of firewalls prefer GUIs to any other interface, while this holds true for only $54$\% of firewall experts.

\subsection{Qualitative Data}\label{subsec:qual}
\begin{figure*}
 \centering
 \includegraphics[width=\textwidth]{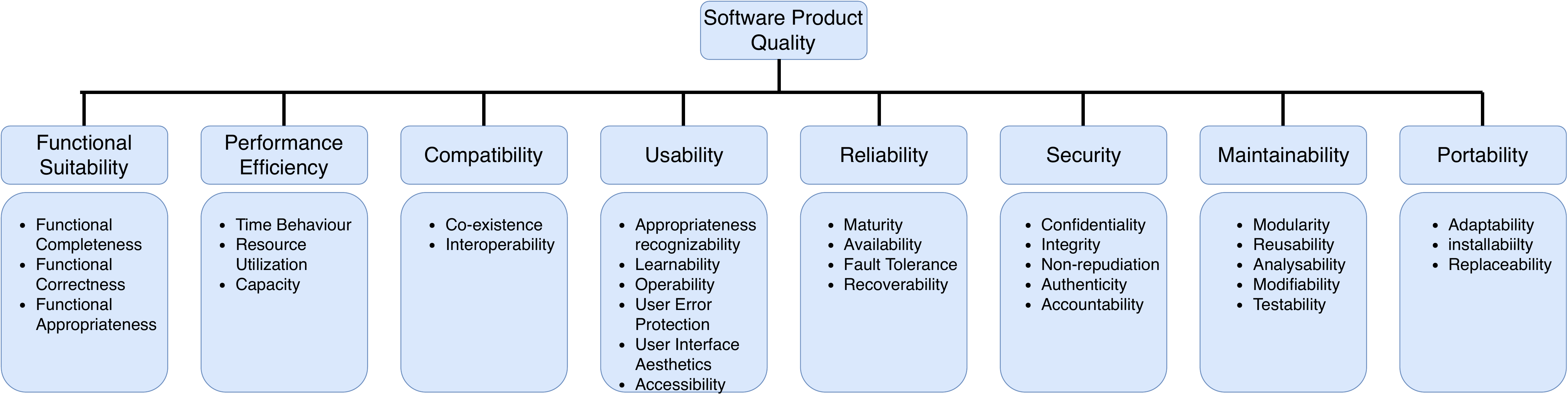}
 \caption{The software quality model ISO/IEC 25010 \cite{organizacion2011iso}.}
 \label{fig:iso}
\end{figure*}

The thoughts and opinions of the system administrators received during our online survey were coded and grouped according to the following principles: 1) the type of interface: CLI, GUI, API, or other; and 2) the type of comment: positive, negative, or neutral. For the convenience of presenting the strengths and limitations of the interfaces, we categorized the codes as follows:
\begin{itemize}
\item We began classifying our codes according to the 10 usability heuristics introduced by Nielsen \cite{nielsen1994usability} (see \mbox{Table \ref{tab:uh}}).
\item Because not all codes concerned usability, some of them did not fall into any of the 10 categories, and were further classified according to the ISO/IEC 25010 \cite{organizacion2011iso}, a standard that defines systems and software quality models (see Figure \ref{fig:iso}). This includes aspects that are not covered by Nielsen's usability heuristics, such as security and reliability. Regarding usability, the ISO standard comprises appropriateness recognizability, learnability, operability, and accessibility, aspects that are not covered by Nielsen's usability heuristics.
\item All remaining codes fell within the \texttt{Other} category. 
\end{itemize}

\begin{table*}[]
\centering
\caption{Nielsen's usability heuristics \cite{nielsen1994usability}.}
\label{tab:uh}
\resizebox{\textwidth}{!}{%
\begin{tabu}{|l|l|}
\hline
\textbf{Heuristics} & \textbf{Short explanation} \\ \hline
Visibility of system status & \begin{tabular}[c]{@{}l@{}}The system
should always keep users informed about what is going on.\end{tabular}
\\ \hline
Match between system and real world & \begin{tabular}[c]{@{}l@{}}The
system should speak the users' language. Information should appear in a
natural\\and logical order.\end{tabular} \\ \hline
User control and freedom & \begin{tabular}[c]{@{}l@{}}Users need
clearly marked emergency exits. The system should support undo and
redo.\end{tabular} \\ \hline
Consistency and standards & \begin{tabular}[c]{@{}l@{}}Users should know
whether different words, situations, or actions mean the same thing.\\
The system should follow platform conventions.\end{tabular} \\ \hline
Error prevention & \begin{tabular}[c]{@{}l@{}}The system should
eliminate error-prone conditions or check for them and present
users\\with a confirmation option before they commit to the
action.\end{tabular} \\ \hline
Recognition rather than recall & \begin{tabular}[c]{@{}l@{}}The system
should minimize the user's memory load. Instructions for use of the
system\\should be visible or easily retrievable whenever
appropriate.\end{tabular} \\ \hline
Flexibility and efficiency of use & \begin{tabular}[c]{@{}l@{}}The
system should have accelerators that can speed up interactions for
expert\\ users so that it can cater to both inexperienced and experienced
users.\end{tabular} \\ \hline
Aesthetic and minimalist design & \begin{tabular}[c]{@{}l@{}}Dialogues
should not contain information that is irrelevant or rarely
needed.\end{tabular} \\ \hline
\begin{tabular}[c]{@{}l@{}}Help users recognize, diagnose, and recover
from\\ errors (we refer to this as assistance with errors)\end{tabular} &
\begin{tabular}[c]{@{}l@{}}The system should explain error messages in
plain language, precisely indicate the\\problem, and constructively
suggest a solution.\end{tabular} \\ \hline
Help and documentation & \begin{tabular}[c]{@{}l@{}}Any system
information should be easy to search, focused on the user's task, list
concrete\\steps to be carried out, and not be too large.\end{tabular} \\
\hline
\end{tabu}%
}
\end{table*}

Because the number of respondents who work with APIs or other interfaces is relatively small, we do not report the corresponding results in this paper. 

\begin{figure}[t]
	\centering
	\includegraphics[width=\columnwidth]{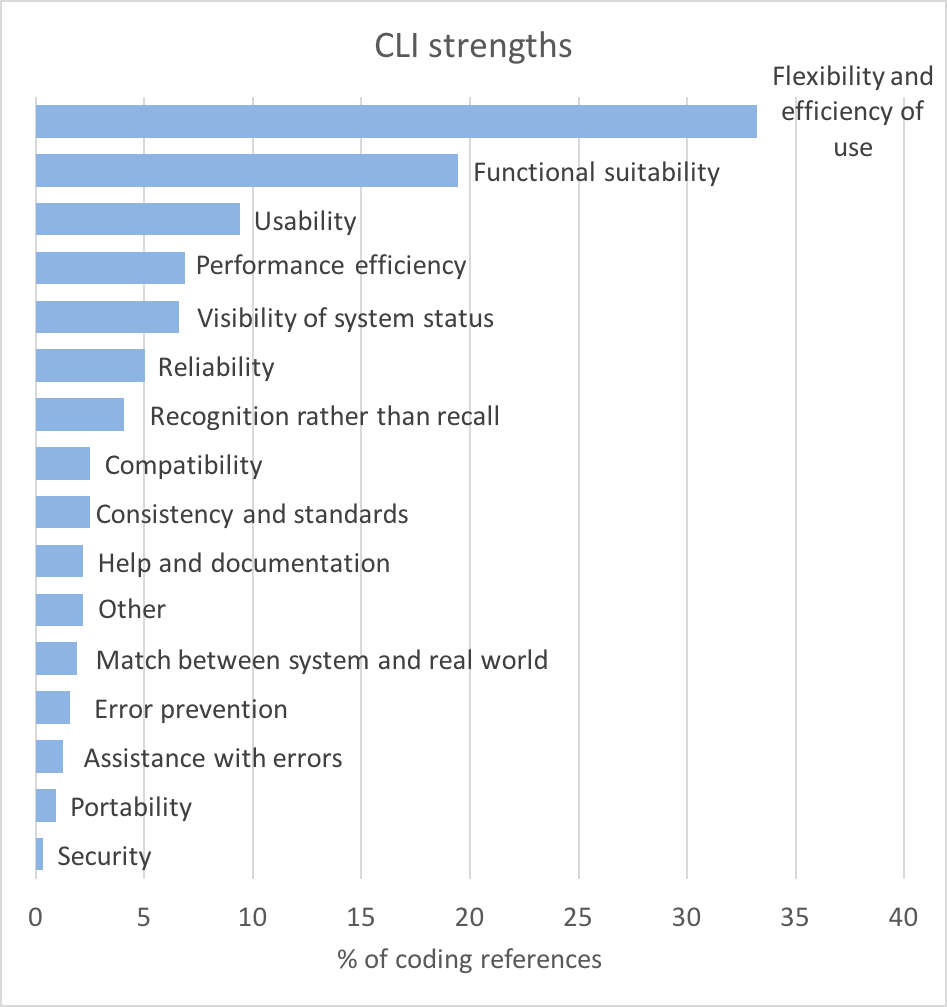}
	\caption{Classification of CLI strengths mentioned by our respondents. The total number of coding references is 319. }
	\label{fig:cli+}
\end{figure}

\begin{figure}[t]
	\centering
	\includegraphics[width=\columnwidth]{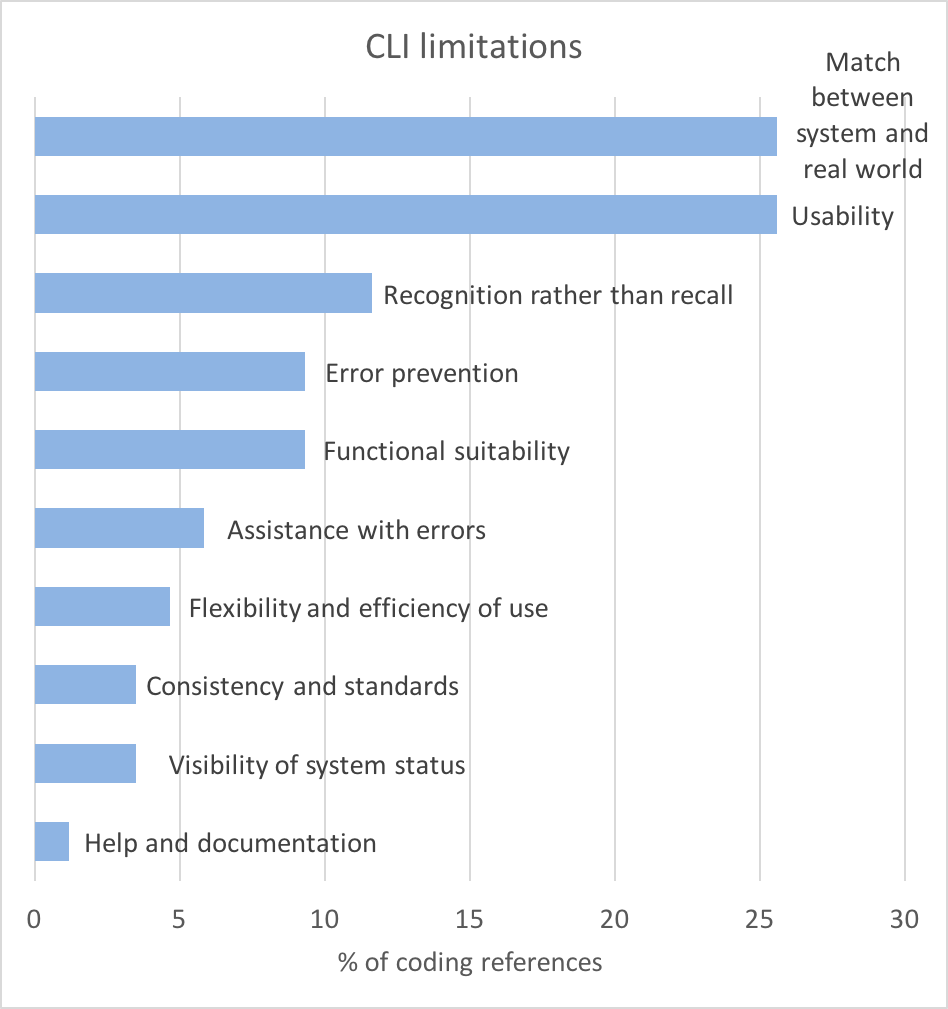}
	\caption{Classification of CLI limitations mentioned by our respondents. The total number of coding references is 86. }
	\label{fig:cli-}
\end{figure}

\begin{figure}[t]
 \centering
 \includegraphics[width=\columnwidth]{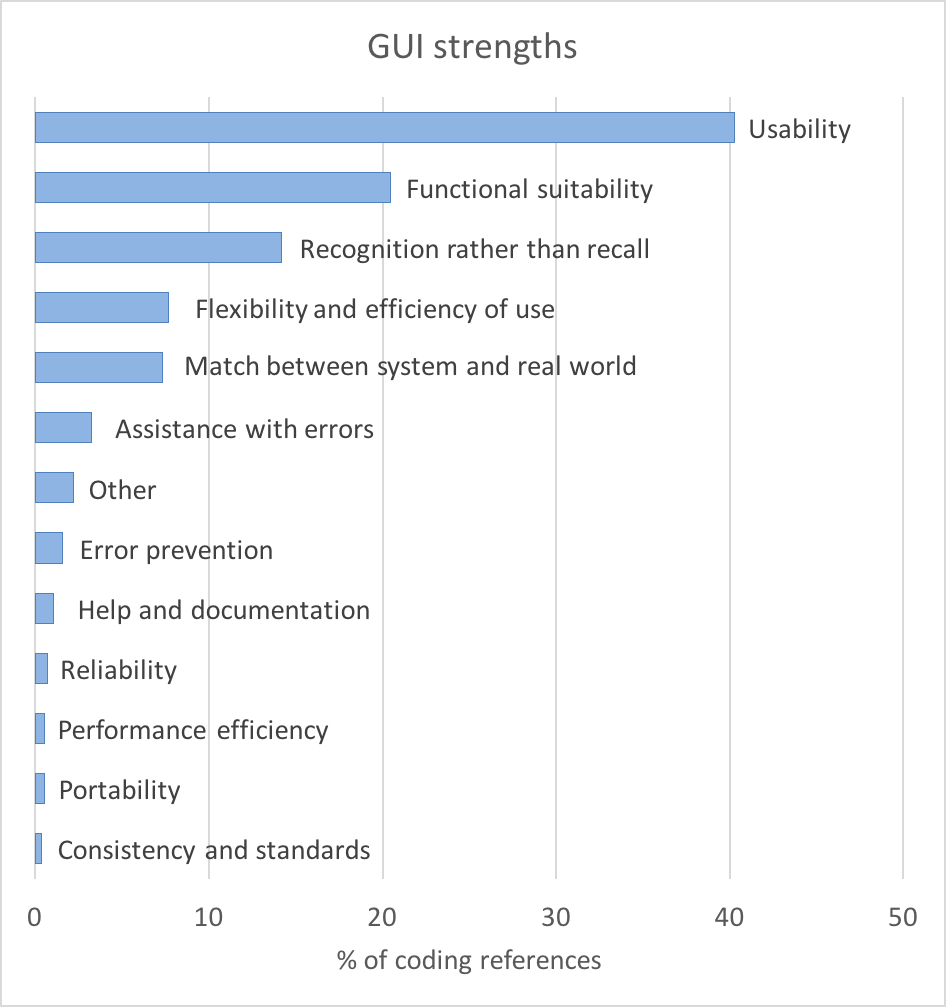}
 \caption{Classification of GUI strengths mentioned by our respondents. Total number of coding references is 586. }
 \label{fig:gui+}
\end{figure}

\begin{figure}[t]
 \centering
 \includegraphics[width=\columnwidth]{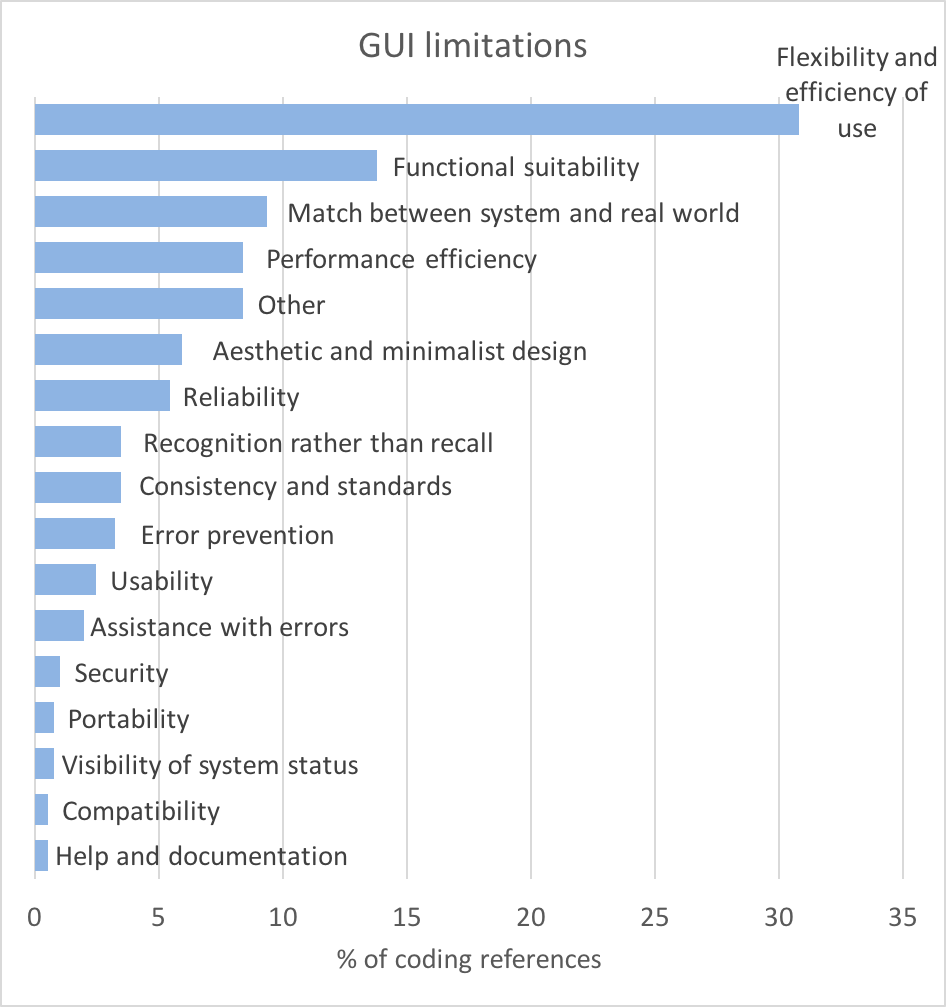}
 \caption{Classification of GUI limitations mentioned by our respondents. Total number of coding references is 406. }
 \label{fig:gui-}
\end{figure}

The strengths and limitations of CLIs and GUIs (see \mbox{Figures \ref{fig:cli+}--\ref{fig:gui-}}) are discussed in further detail in \mbox{Sections \ref{subsec:cli-str}--\ref{subsec:gui-lim}}. In each subsection, we examine the categories that cover 80\% of all coding references, starting from the most popular. Note that subsections have different total numbers of coding references, and not all codes in each category are discussed in detail. For convenience, codes are highlighted in bold.

\subsubsection{CLI Strengths}\label{subsec:cli-str}
According to our respondents, CLIs have a number of strengths (the total number of coding references is 319):
\begin{enumerate}
\item Flexibility and efficiency of use ($106$ coding references; $33.2$\%). Several respondents (64 coding references) noted the possibility of \textbf{automation} as a strength of CLIs: \emph{``CLIs are good targets for automation, even if the only thing you can do is bash scripting.''} \textbf{User efficiency} was mentioned 42 times. One respondent stated: \emph{``CLIs have a high signal-to-noise ratio, and are therefore preferable to everything else.''}
\item Functional suitability ($62$ coding references; $19.4$\%). The \textbf{superior functionality} of CLIs was mentioned 37 times by system administrators: \emph{``100\% coverage of all firewall functionality supported by the OS kernel, unlike GUIs and APIs.''} Other useful features of the interface, such as the \textbf{ability to work offline} and \textbf{ease of search}, were stated 16 times. 
\item Usability ($30$ coding references; $9.4$\%). According to 12 system administrators, the user has \textbf{full control} with a firewall CLI: \emph{``I do not see any reasonable way to be sure a firewall is doing the right thing without using a CLI.''} Seven other respondents stated the advantages of managing a firewall with a CLI: \emph{``Properly used, CLI is by far the best method to manage any system.''}
\item Performance efficiency ($22$ coding references; $6.9$\%). The system administrators noted the superior \textbf{speed of operation} of CLIs (22 coding references), commenting \emph{``[CLI] uses zero system resources''} and \emph{``it is faster and does not take five minutes to load.''}
\item Visibility of system status ($21$ coding references; $6.6$\%). \textbf{Transparency} was mentioned 21 times as an important positive characteristic of CLIs: \emph{``With a CLI, you know exactly what the firewall is doing.''}
\item Reliability ($16$ coding references; $5.0$\%). Our respondents highlighted some strengths of CLIs, such as: \textbf{reliability}: \emph{``... there is a lower incidence of random issues with the UI''}; \textbf{high availability}: \emph{``I can do the same task via an SSH connection or even a KVM if the whole network is down. I can do that via a smartphone if I must.''}; and ease of \textbf{configuration backup}: \emph{``Backing up and restoring configurations easily through text files.''}
\end{enumerate}

\subsubsection{CLI Limitations}\label{subsec:cli-lim}
The main CLI limitations noted by our respondents are the following (the total number of coding references is 86):
\begin{enumerate}
\item Match between system and real world ($22$ coding references; $25.6$\%). The main problem, which was referenced 19 times, is a \textbf{long learning curve}. System administrators shared that \emph{``CLI may be scary/overwhelming for a beginner/untrained user''} and \emph{``There is typically a slightly higher learning curve associated with CLI, which can often be discouraging to unexperienced users.''}
\item Usability ($22$ coding references; $25.6$\%). There are two codes that were referenced more than any others: CLIs are \textbf{not easy to use} (8 times) and \textbf{inconvenient data representation} (7 times). Two respondents stated: \emph{``The CLI is not capable of representing all the firewall rule data in a clean and easy-to-read format''} and \emph{``CLIs are terrible at generating visual information that is comprehensible by non-experts...''} Regarding the ease of use, one system administrator wrote that \emph{``ease of use is a definite issue [of CLI].''}
\item Recognition rather than recall ($10$ coding references; $11.7$\%). The facts that CLIs are \textbf{less intuitive} and \textbf{less educational} were mentioned seven and three times, respectively. CLIs \emph{``may be less intuitive than other interfaces''} and \emph{``you cannot click your way around it in an attempt to \emph{figure it out}.''}
\item Error prevention ($8$ coding references; $9.3$\%). CLIs are \textbf{prone to errors}, both typographical and logical, and that fact was named 8 times by the respondents. One system administrator wrote that it is \emph{``much easier to cause catastrophic failure quickly and effectively''} with a CLI.
\item Functional suitability ($8$ coding references; $9.3$\%). The absence of some auxiliary functionality was noted by eight respondents. A CLI \emph{``has no Ctrl+F [searching] feature.''}
\end{enumerate}

\subsubsection{GUI Strengths}\label{subsec:gui-str}
GUIs have several strengths (the total number of coding references is 586):
\begin{enumerate}
\item Usability ($236$ coding references; $40.3$\%). In general, GUIs are known to be user-friendly. Visual representations of data provide a \textbf{better understanding and/or overview of configuration} according to 124 coding references. One respondent shared with us that \emph{``it [GUI] allows me to have a better understanding of a firewall's configuration while having that information displayed in a more organized manner when compared to a CLI.''} The system administrators also stated that GUIs are \textbf{easy to use} (49 times), \textbf{easy to manage and modify rules} with (19 times), good for \textbf{creating rules and policies} (16 times) and \textbf{good for people that struggle to work with text} (six times).
\item Functional suitability ($120$ coding references; $20.5$\%). The system administrators wrote that GUIs are excellent for a variety of tasks, such as \textbf{monitoring} (17 coding references), \textbf{reporting} (nine coding references), and \textbf{logging} (five coding references). Another strong aspect of GUIs is an \textbf{ease of displaying additional information} (20 coding references), such as graphs and statistics. 
\item Recognition rather than recall ($83$ coding references; $14.2$\%). Being easy to navigate, GUIs are an irreplaceable tool that is \textbf{good for occasional use} (44 coding references). A system administrator shared: \emph{``Because of my responsibilities as a general sysadmin [system administrator], management of the firewall takes up only a small part of my time, and having using the GUI for management means that I do not have to remember CLI commands.''}
\item Flexibility and efficiency of use ($45$ coding references; $7.7$\%). \textbf{User efficiency} was named 20 times as a strength of GUIs: \emph{``Makes it faster than using CLI to edit basic things on a firewall \ldots,''} \emph{``It just gives me a quicker and more visual grasp on what I am doing. Point, click, move on...''}
\end{enumerate}

\subsubsection{GUI Limitations}\label{subsec:gui-lim}

Although the majority of our respondents prefer to use GUIs to other alternatives, several serious limitations of GUIs were named (the total number of coding references is 406):
\begin{enumerate}
\item Flexibility and efficiency of use ($125$ coding references; $30.8$\%). A \textbf{lack of automation} and \textbf{user inefficiency} in general when working with the interface were stated 102 times. This makes GUIs less useful for experts. One system administrator wrote: \emph{``We are at a very bad time for GUI firewalls, because experts are the only ones who can effectively scale the workloads demanded of the modern IT infrastructure, and GUIs are almost useless for most experts in that regard.''}
\item Functional suitability ($56$ coding references; $13.8$\%). According to 56 participants, the \textbf{reduced functionality} of GUIs is a serious issue: \emph{``[GUI is] missing a lot of features/settings, so that you have to use CLI to make changes.''}
\item Matching between the system and real world ($38$ coding references; $9.4$\%). Because a GUIs represents an \textbf{additional layer of abstraction}, the user may lack a deeper understanding of their actions (30 coding references). A system administrator formulated a drawback of GUIs as \emph{``a lack of knowledge for the underlying system you are working on.''} Another problem, named six times, is that GUIs may \textbf{generate less understandable configuration files}: \emph{``GUIs do not always generate configs that make logical/visual sense to a human.''}
\item Performance efficiency ($34$ coding references; $8.4$\%). GUIs are highly demanding in terms of system resources, and for this reason are usually very \textbf{slow} (34 references): \emph{``GUIs take more overhead to display and run, which may draw away from a firewall's processing power.''}
\item Other ($34$ coding references; $8.4$\%). The system administrators stated a number of problems. The facts that GUIs \textbf{require additional equipment or software} and are \textbf{platform or browser dependent} were mentioned 12 times each. Two participants shared that \emph{``A software client can be needed, which may not always be accessible...''} and \emph{``Depending on browser it can be a horrible experience (slow, unresponsive, thus can cause issues with clicks being registered late or not at all).''} Several additional issues were mentioned by the system administrators, such as GUIs being \textbf{difficult to document} (five references): \emph{``Unlike CLIs, documenting a GUI is mostly useless and defeats most of the purpose of a GUI,''} and \textbf{unavailable for particular firewalls} (three references): \emph{``I currently do not have a firewall that supports \mbox{GUI \ldots}.''} 
\item Aesthetic and minimalist design ($24$ coding references; $5.9$\%). The respondents encountered \textbf{badly designed} GUIs (16 references): \emph{``\ldots some [GUIs] are horribly designed so it is hard to figure out how to do what you want to do.''} Eight system administrators noted the problem of an \textbf{interface beauty and functionality trade-off}: \emph{``Most [G]UIs are either poorly laid out making them difficult to use or are too user-friendly and do not have all settings available.''}
\item Reliability ($22$ coding references; $5.4$\%). The respondents mentioned reliability issues (22 times) with GUIs: \emph{``They can crash and become unresponsive, sometimes you cannot tell if it is processing a new config/update or if it is locked up.''} 
\end{enumerate}

\subsection{Suitability for Different Tasks}\label{subsec:suit}

In addition to the strengths and limitations of firewall CLIs and GUIs, the system administrators informed us of which interface they deem to be the most suitable for each task. 

All use cases associated with entering many similar rules on one computer or bulk changes on several computers at once require the use of a CLI. Furthermore, non-standard tasks in which, for example, rules with a set of advanced options are necessary, are more easily solved using a CLI according to the respondents. One system administrator stated:
\begin{Quote}
\emph{GUIs and APIs are terrible at handling special cases and rarely expose the underlying command structure properly (they always create a monopoly on how things are done). With CLIs, it is usually a straightforward process to use the underlying commands and kernel modules directly (iptables and netfilter on Linux).}
\end{Quote}

Firewall GUIs are preferable in more tasks according to the polled system administrators. The most frequently mentioned use cases in which the use of a GUI is beneficial are the creation of individual rules and building entire configurations. One respondent commented: 
\begin{Quote}
\emph{The more complicated a task is, the more important a GUI becomes.  Who wants to set complex web proxy configuration options via CLI?  Let us say that I want to proxy students, teachers, and administrators in a school differently.  Let us start with just http request options for request headers, allowed auth[entication] methods, DLP scanning for HTTP POST, etc.  Imagine the difference between clicking checkboxes and dropdown menus vs trying to type all this out via a CLI with some reference \hbox{manual \ldots}}
\end{Quote}
Another task that is easier to perform in a GUI is viewing and inspecting firewall rules and policies. GUIs usually have an option to link objects with rules, which allows the bigger picture to be observed. 
\begin{Quote}
\emph{Examining and working with firewall rules is an instructive example of where the CLI is not a good option.  The CLI is not capable of representing all the firewall rule data in a clean and easy-to-read format. This is much easier on a GUI.}
\end{Quote}
Monitoring is another use case in which the system administrators decide to use GUIs. GUIs provide the ability to view connection statistics, monitor traffic flows through real-time graphs, and so on, and are therefore preferred. The system administrators also tend to choose GUIs when there is a need to change the order of rules in a rule set.

%% file: tex/discussion.tex
\section{Discussion}\label{sec:discussion}

The collected quantitative data provides insights into the usage of different firewall interfaces that are considerably different from what has been previously published in the literature. We observe a significant shift towards the use of GUIs, although CLIs have been widely utilized by system administrators in the recent past \cite{takayama2006trust}. There are three possible explanations for this shift. 

The first concerns the case of security tools, in particular firewalls, where designers attempt to follow the design principles formulated for system administration tools. Our participants confirmed that firewall GUI implementations are improving. One system administrator opined: 
\begin{Quote}
\emph{Decades of GUI development: a 2D mouse and keyboard with a keyboard shortcuts interface instead of the serial text in/out of a classic CLI. More available and powerful searching, sorting, and filtering of information; discoverability of available commands; and visual/graphical possibilities of a large and high-res screen.}
\end{Quote}

The second possible reason is that the number of system administrators has significantly increased, including those with limited technical expertise, as described by Xu et al. \cite{xu2016hci}. The statement on less experienced system administrators is not valid for our data sample, as $83$\% of the respondents have worked for over three years as system administrators (see Table \ref{tab:demographics}). 

The third possible reason is that there are many system administrators who are not security experts, but rather general purpose system administrators.
As we can also observe from Table \ref{tab:demographics}, $74$\% of the respondents spend no more than four hours per week managing firewalls. Therefore, they are most likely general purpose system administrators, and this explains their reluctance to work with firewall CLIs, which are less usable, require more learning time, and are prone to errors according to our participants. 

Our qualitative data show that GUIs are less preferable for experts compared to system administrators with a lower firewall proficiency. The respondents noted that GUIs are not very useful for experts, as they severely restrict the user with limited functionality, a low operation speed, and low user interaction efficiency owing to the lack of automation capabilities.

Another feature that we noticed when analyzing the data from the survey is that our respondents' preferences for one interface do not always depend on their strengths or limitations. Sometimes system administrators are more comfortable with a CLI or GUI simply because they familiarized themselves with this interface first. One respondent stated: 
\begin{Quote}
\emph{I am old school and there were no GUIs back in the day, so it [CLI] is more comfortable for me.} 
\end{Quote}
Another possible reason is that system administrators do not always have experience with other interfaces, and therefore cannot objectively compare their strengths and limitations.

\subsection{Limitations}
One of the limitations of our study is that most of the respondents were recruited through online forums for system administrators. Because the survey participants were volunteers, there is a self-selection bias that leads to the sample not being fully representative. 

Furthermore, the study was conducted online and we could not observe the participants answering the questions. Moreover, some of the answers were ambiguous, and so we had to interpret them, which could lead to a distortion of the meaning that the respondent had originally intended. For example, the comment \emph{``slow''} can refer both to the speed of operation of the software and the speed of interaction between the user and interface. There is also a possibility of questions being misunderstood or misinterpreted by the participants. Additionally, self-report surveys have several common limitations \cite{gonyea2005self}, such as social desirability biases and acquiescent responses.

We mitigated the limitations by carefully considering the design of the survey, pretesting it with several participants, making it anonymous so that people could answer honestly, and shortening it to minimize respondent fatigue.

\subsection{Design Recommendations}
Our survey identified some problems for both CLIs and GUIs that should be taken into account. In this section, we present some design recommendations for CLIs and GUIs based on the results of our survey, as well as discussing the benefits of combining these two interfaces into one. 

As one respondent noted: 
\begin{Quote}
\emph{CLI interfaces are not usually as forgiving as other interfaces. If you are not paying attention, then the slightest typo could cause large issues.}
\end{Quote}
Our recommendation is to employ a syntactical verification of commands when a user types in an instruction to prevent errors in firewall configuration processes. 

Furthermore, because CLIs have a reasonably long learning curve, assistance in writing rules is necessary for less experienced system administrators. Respondents noted the following:
\begin{Quote}
\emph{It may be difficult to compose rules [in CLIs] without an example.}
\end{Quote}
Providing a knowledge base of examples of rules could be a useful approach.

We make three recommendations regarding GUIs. First, the system administrators complained about the speed of operation of GUIs. Our recommendation is to not make GUIs bloated, so that they do not consume a lot of system resources and can be run on mediocre hardware.

The second recommendation relates to the GUI installation process. As one of the system administrators commented: 
\begin{Quote}
\emph{They [GUIs] are not really for beginners because of the initial setup required to configure them.}
\end{Quote}
Because the highest percentage of GUI use is among system administrators with the least firewall expertise, installing a firewall should not be a complicated task. 

In addition, to increase the speed of user interaction with a firewall GUI it is necessary to allow system administrators to create their own combinations of hotkeys for the most popular actions. This will help to make GUIs more attractive for firewall experts.

While we have provided recommendations for how to improve each interface, there remain problems that are difficult to solve within one interface. For example, textual interfaces are inherently inadequate for presenting a large amount of information: 
\begin{Quote}
\emph{When a config file has over 2000+ lines it is easy to lose track of what is what [in CLI].} 
\end{Quote}
Another limitation originates from the concept of a CLI: it is impossible to create and edit rules using the mouse cursor and check boxes, which in some cases can significantly increase the productivity of a firewall operator. For GUIs, the problem is the lack of automation tools, as was noted by a large number of respondents. 

A more effective solution would be to combine two interfaces into one, with the ability to seamlessly switch from one to the other, so that interacting with one interface affects the other. Such an approach can leverage the strengths of each interface while mitigating their limitations \cite{murillo2014empowering}. A GUI can provide an overview of configurations and display additional graphs and statistics, as well as being used to create rules, while a CLI can offer on-demand access to the powerful automation capabilities. Such a combined interface could be suitable for users with different firewall expertise. Less experienced system administrators could be trained to use the CLI by viewing the underlying text-based commands while working in the GUI. Expert users could continue using commands to create rules, while using the GUI for a better policy overview. We strongly believe that such a firewall interface would be widely accepted in the system administration community.

%% file: tex/conclusion.tex
\section{Conclusion}\label{sec:conclusion}

In this work, we present an online study concerning system administrators, in which we examine how they interact with different firewall interfaces. The survey results show that $70$\% of the polled system administrators are primarily GUI users, and $60$\% prefer this interface for interacting with a firewall. This finding differs from previously published findings in the literature, in which CLIs were claimed to be the first choice of system administrators.  

We classify the strengths and limitations of firewall CLIs and GUIs. Our participants report that CLIs are flexible, efficient, transparent, reliable, and achieve ultimate functionality and a good performance. However, they are inconvenient for representing data, do not help users by preventing errors, and have a long learning curve. On the other hand, GUIs help users to perceive firewall configuration information more effectively and have a shorter learning curve compared to CLIs. They are also easy to use, easy to create and modify rules with, and good for occasional use. Regarding the limitations of GUIs, they restrict users with limited functionality, a low operation speed, and a low user interaction efficiency. They are neither transparent nor reliable. In addition, we report the preferred interface for each task according to the system administrators.

Our findings present opportunities for future research. A well-designed firewall interface should predict and interpret its user's needs and assist them in becoming proficient with the firewall. In this case, the system administrator is satisfied with the firewall and can efficiently perform the required work. On the other hand, a poorly designed firewall interface might hinder the successful execution of tasks and lead to the future disuse of that solution. We provide some design recommendations that should be taken into account by designers aiming to develop better CLIs and GUIs.

%% file: tex/acknowledgment.tex
\section*{Acknowledgments}
We are grateful to all the system administrators that participated in our study. We would also like to thank the moderators of the \emph{Sysadmin} subreddit (\texttt{r/sysadmin}) for allowing us to reach out to their community.

This work was supported by the Knowledge Foundation of Sweden {\small HITS} project and by the Swedish Foundation for Strategic Research {\small SURPRISE} project.

%% file: tex/availability.tex
\section*{Availability}\label{sec:availability}

The final codes and other details are available at \url{https://github.com/soups2019-126/supplementary_material}.

%% file: tex/appendix.tex
\begin{appendices}
\renewcommand{\labelitemi}{$\circ$}

\section{Survey Questions}
\begin{center}
\textbf{Page 1}
\end{center}
\begin{enumerate}
\item How much time per week (on average) do you spend directly interacting with (managing) firewalls?
	\begin{itemize}
	\item Less than 1 hour/week
	\item 1--4 hours/week
	\item 5--8 hours/week
	\item 9--12 hours/week
	\item More than 12 hours/week
	\item I do not directly manage firewalls
	\end{itemize}
\item Can you enumerate all the firewall-related tasks that you have dealt with?
	\begin{itemize}
	\item Adding/removing firewall rules
	\item Examining firewall policies to understand their functionalities
	\item Inspecting firewall rules/policies to find errors or inconsistencies 
	\item Other (please specify)
	\end{itemize}
\item What is the PRIMARY firewall interface that you use at work?\footnote{\%\emph{primary}\% returns the selected option in Question 3}
	\begin{itemize}
	\item Command Line Interface (CLI)
	\item Graphical User Interface (GUI)
	\item Application Programming Interface (API)
	\item Other (please specify)
	\end{itemize}
\bigskip
\vspace*{1cm}
\item What is your PREFERRED firewall interface?\footnote{\%\emph{preferred}\% returns the selected option in Question 4}
	\begin{itemize}
	\item Command Line Interface (CLI)
	\item Graphical User Interface (GUI)
	\item Application Programming Interface (API)
	\item Other (please specify)
	\end{itemize}
\end{enumerate}

\begin{algorithmic}
\If {\emph{answer(Q3)} = \emph{answer(Q4)}}
	\State \textbf{go to} Page 2
\ElsIf{\emph{answer(Q3)} = 2}
	\State \textbf{go to} Page 3
\ElsIf{\emph{answer(Q4)} = 2}
	\State \textbf{go to} Page 4
\Else
	\State \textbf{go to} Page 5
\EndIf
\end{algorithmic}

\smallskip
\begin{center}
\textbf{Page 2}
\end{center}
\begin{enumerate} [resume]
\item Are there certain tasks that the \%\emph{preferred}\% allows you to do, which are more difficult to do using other firewall interfaces?
\item What are the strengths of the \%\emph{preferred}\%, if any?
\item Can you think of any problems associated with the \%\emph{preferred}\%?
\end{enumerate}

\begin{algorithmic}
\If {\emph{answer(Q3)} = 2}
	\State \textbf{go to} Page 10
\Else
	\State \textbf{go to} Page 5
\EndIf
\end{algorithmic}

\bigskip
\begin{center}
\textbf{Page 3}
\end{center}
\begin{enumerate} [resume]
\item Why do you prefer the \%\emph{preferred}\% to the \%\emph{primary}\% when managing firewalls?
\item What are the strengths of the \%\emph{preferred}\%, if any?
\item Do you see any strengths in the \%\emph{primary}\%?
\item What problems do you see with the \%\emph{primary}\%?
\end{enumerate}

\begin{algorithmic}
\State \textbf{go to} Page 10
\end{algorithmic}

\bigskip
\begin{center}
\textbf{Page 4}
\end{center}
\begin{enumerate} [resume]
\item Why do you prefer the \%\emph{preferred}\% to the \%\emph{primary}\% when managing firewalls?
\item What are the strengths of the \%\emph{preferred}\%, if any?
\item Can you think of any problems associated with the \%\emph{preferred}\%?
\item Do you see any strengths in the \%\emph{primary}\%?
\end{enumerate}

\begin{algorithmic}
\State \textbf{go to} Page 10
\end{algorithmic}

\bigskip
\begin{center}
\textbf{Page 5}
\end{center}
\begin{enumerate} [resume]
\item Have you ever used a graphical user interface (GUI) to manage a firewall?
	\begin{itemize}
	\item Yes
	\item No
	\end{itemize}
\end{enumerate}

\begin{algorithmic}
\If {\emph{answer(Q16)} = 2}
	\State \textbf{go to} Page 6
\Else
	\State \textbf{go to} Page 7
\EndIf
\end{algorithmic}

\bigskip
\begin{center}
\textbf{Page 6}
\end{center}
\begin{enumerate} [resume]
\item Can you name the reasons for not trying a firewall graphical user interface (GUI)?
\end{enumerate}

\begin{algorithmic}
\State \textbf{go to} Page 10
\end{algorithmic}

\bigskip
\begin{center}
\textbf{Page 7}
\end{center}
\begin{enumerate} [resume]
\item Are you currently using a GUI to manage your firewall?
	\begin{itemize}
	\item Yes
	\item No
	\end{itemize}
\end{enumerate}

\begin{algorithmic}
\If {\emph{answer(Q18)} = 2}
	\State \textbf{go to} Page 8
\Else
	\State \textbf{go to} Page 9
\EndIf
\end{algorithmic}

\bigskip
\begin{center}
\textbf{Page 8}
\end{center}
\begin{enumerate} [resume]
\item Can you name the reasons for not using a firewall with a GUI and whether you see problems with GUIs?
\end{enumerate}

\begin{algorithmic}
\State \textbf{go to} Page 10
\end{algorithmic}

\bigskip
\begin{center}
\textbf{Page 9}
\end{center}
\begin{enumerate} [resume]
\item For which tasks do you use the firewall graphical user interface (GUI)?
\end{enumerate}

\bigskip
\begin{center}
\textbf{Page 10}
\end{center}
\begin{enumerate} [resume]
\item How long have you been working as a system/network administrator?
	\begin{itemize}
	\item Less than a year
	\item 1--3 years	
	\item 4--6 years
	\item 7--9 years
	\item 10 years and more
	\end{itemize}
\item How would you describe your proficiency with firewalls?
	\begin{itemize}
	\item Basic knowledge
	\item Intermediate	
	\item Advanced
	\item Expert
	\end{itemize}
\item How old are you?
	\begin{itemize}
	\item 18--24 years old
	\item 25--34 years old	
	\item 35--44 years old
	\item 45--54 years old
	\item 55--64 years old
	\item 65 years or older
	\item Prefer not to answer
	\end{itemize}
\item What is your gender?
	\begin{itemize}
	\item Female
	\item Male	
	\item Other
	\item Prefer not to answer
	\end{itemize}
\end{enumerate}

\end{appendices}